\documentstyle[preprint,aps,epsf,floats]{revtex}

\begin{document}

\draft

\title{
Radiative corrections to all charge assignments of heavy quark baryon
semileptonic decays}

\author{
A.\ Mart{\'\i}nez
}
\address{
Escuela Superior de F\'{\i}sica y Matem\'aticas del IPN, \\
Apartado Postal 75-702, M\'exico, D.F.\ 07738, Mexico
}

\author{
J.\ J.\ Torres
}

\address{
Escuela Superior de C\'omputo del IPN, \\
Apartado Postal 75-702, M\'exico, D.F. 07738, Mexico
}

\author{A.\ Garc{\'\i}a
}

\address{
Departamento de F{\'\i}sica, Centro de Investigaci\'on y de Estudios Avanzados del IPN \\
Apartado Postal 14-740, M\'exico, D.F.\ 07000, Mexico
}

\author{
Rub\'en Flores-Mendieta
}

\address{
Instituto de F{\'\i}sica, Universidad Aut\'onoma de San Luis Potos{\'\i}, \\
\'Alvaro Obreg\'on 64, Zona Centro, San Luis Potos{\'\i}, S.L.P.\ 78000, Mexico
}

\date{June 29, 2002}

\maketitle

\tightenlines

\begin{abstract}
In semileptonic decays of spin-1/2 baryons containing heavy quarks up to six
charge assignments for the baryons and lepton are possible. We show that the
radiative corrections to four of these possibilities can be directly
obtained from the final results of the two possibilities previously studied.
There is no need to recalculate integrals over virtual or real photon
momentum or any traces.
\end{abstract}


\section{Introduction}

By necessity, the calculation of radiative corrections (RC) to spin-1/2
baryon semileptonic decays (BSD) requires that definite charges be chosen
for the participating particles. In previous calculations~\cite{c1} we have chosen
a negatively charged emitted lepton and either a neutral or a negatively
charged decaying baryon. This covers all of hyperon semileptonic decays,
with the exception of $\Sigma^{+}\rightarrow \Lambda l^{+}\nu_l$. 
If the polarization of the baryons is not involved, the previous results can
be extended~\cite{c2} to cover this latter decay using a very practical rule, which
also applies to $\Lambda_{c}^{+}\rightarrow \Lambda l^{+}\nu_l$.
However, when heavy quarks are involved several other charge arrangements
for the baryons appear, and then one faces the problem of having to
recalculate the accompanying RC. In particular doubly-charged baryons must
also be considered. Recently, preliminary evidence for double-charm baryons
has been reported~\cite{c3}, so it is also timely to study the RC to semileptonic
decays of such baryons. It is the purpose of this paper to obtain the RC to
BSD with all the charge assignments to the baryons allowed when heavy quarks
are involved. Our main result will show that with proper adaptations, the
previous results obtained with the initial choices of baryon charges can
also be used to obtain the RC to the other charge assignment possibilities.

Let us list the several types of BSD we have to discuss. For definiteness we
shall take as a heavy quark the charm quark, i.e., we shall consider the
four quarks $u$, $d$, $s$, and $c$. Their charges are $2/3$, $-1/3$, $-1/3$, and $2/3$ and
their strong flavors are $S=0$, $0$, $-1$, $0$, $C=0$, $0$, $0$, $1$, and $I_{3}=1/2$, $-1/2$,
$0$, $0$, respectively. Their semileptonic decays are
\begin{eqnarray}
d &\rightarrow & u l^{-} \overline{\nu}_l \qquad \left( \Delta S=\Delta
C=0,\;\Delta Q=1\right), \\
s &\rightarrow &ul^{-}\overline{\nu}_l \qquad \left( \Delta C=0,\;\Delta
S=\Delta Q=1\right), \\
c &\rightarrow &sl^{+} \nu_l \qquad \left( \Delta S=\Delta
C=\Delta Q=-1\right), \\
c &\rightarrow &dl^{+} \nu_l \qquad \left( \Delta S=0,\;\Delta
C=\Delta Q=-1\right).
\end{eqnarray}
In parentheses we display the selection rules these decays obey, in terms of
the charges and strong flavors involved. In addition, the
Cabibbo-Kobayashi-Maskawa matrix elements should accompany these selection
rules.

At the hadron level these quarks can form twenty baryons. Eight carry no
charm, nine carry single charm, and three carry double charm. The
semileptonic decays of these baryons are grouped into
\begin{eqnarray}
&  & A^{-} \rightarrow  B^{0}( l^{-}\overline{\nu}_l ), \\
&  & A^{0} \rightarrow  B^{+}( l^{-}\overline{\nu}_l ), \\
&  & A^{+} \rightarrow  B^{0}( l^{+}\nu_l ), \\
&  & A^{0} \rightarrow  B^{-}( l^{+}\nu_l ), \\
&  & A^{++} \rightarrow  B^{+}( l^{+}\nu_l ), \\
&  & A^{+} \rightarrow B^{++}( l^{-}\overline{\nu }_l ).
\end{eqnarray}

The BSD of the eight hyperons fall in the groups (5) or (6), with the
exception of $\Sigma^{+}\rightarrow \Lambda l^{+}\nu_l$ which
falls in group (7). Many charm baryon (or bottom baryon) decays will fall in
some of the groups (5)-(7), but the last groups (8)-(10) necessarily require
the intervention of charm. For completeness, let us list the BSD with
appreciable phase-space according to the above groups. We use the naming
scheme of Ref.~\cite{c4}, whose Greek symbol indicates isospin and subindex
indicates heavy quark content. All of these decays are driven, at the quark
level, by the semileptonic decays (2), (3), or (4),
\begin{eqnarray}
\begin{array}{lll}
{\rm Group \ (7)}: & \Omega_{cc}^{+}\rightarrow \Omega_c^0 l^+ \nu_l, &
\Omega_{cc}^+ \rightarrow \Xi_c^{\prime \, 0} \left(\Xi_c^0 \right) l^+ \nu_l,  \\
 & \Xi_{cc}^+ \rightarrow \Xi_c^{\prime \, 0} \left(\Xi_c^0 \right) l^+ \nu_l, &
\Xi_{cc}^+ \rightarrow \Sigma_c^0 l^+ \nu_l, \\
 & \Xi_c^+ \rightarrow \Xi^0 l^+ \nu_l,
 & \Xi_c^+ \rightarrow \Lambda \left( \Sigma^0 \right) l^+ \nu_l, \\
 & \Lambda_c^+ \rightarrow \Lambda \left( \Sigma^0 \right) l^+ \nu_l, &
\Lambda_c^+ \rightarrow n l^+ \nu_l. \\
{\rm Group \ (8)}: & \Xi_c^0 \rightarrow \Xi^- l^+ \nu_l, &
\Xi_c^0 \rightarrow \Sigma^- l^+ \nu_l, \\
 & \Omega_c^0 \rightarrow \Sigma^- l^+ \nu_l. \\
{\rm Group \ (9)}: & \Xi_{cc}^{++} \rightarrow \Xi^{\prime +} \left( \Xi_c^+ \right) l^+ \nu_l, &
\Xi_{cc}^{++} \rightarrow \Lambda_c^+ \left( \Sigma_c^+ \right) l^+ \nu_l. \\
{\rm Group \ (10)}: & \Omega_{cc}^+ \rightarrow \Sigma_{cc}^{++} l^- \overline{\nu}_l, &
\Xi_c^+ \rightarrow \Sigma_c^{++} l^- \overline{\nu}_l.
\end{array} \nonumber
\end{eqnarray}

We have omitted above those decays with very small phase space driven by (1)
and also those decays which are overwhelmed by strong decays, like $\Sigma
_{c}\rightarrow \Lambda_{c}^{+}\pi,\;$or by electromagnetic decays, like $%
\Xi_{c}^{\prime \, 0}\rightarrow \Xi_{c}^{0}\gamma$.

In what follows we shall obtain model-independent RC, according to the
analyses of Refs.~\cite{c5,c6}, which include terms of zeroth and first order
in $\left( \alpha /\pi \right) \left( q/M_{1}\right)^{0}$ and $\left(
\alpha /\pi \right) \left( q/M_{1}\right),$ where $q$ is the four-momentum
transfer and $M_{1}$ is the mass of $A.$ We shall not impose any kinematical
constraint on the four-momentum of the bremsstrahlung photon, so that our
results will be useful both in what in previous papers we referred to as the
three-body and four-body regions. We shall allow for non-zero polarization $s_1$
of the decaying baryon $A$. Our first task is to extend the
RC with polarized $A$ obtained for $l^{-}$ emission in groups (5) and
(6), to decays with $l^{+}$ emission in groups (7) and (8). This is done
in Sect.~II. A simple and practical rule is obtained, analogous to the rule
of Ref.~\cite{c2}. The groups of decays involving double charge of one of the
baryons will be discussed in Sect.~III. Although this requires more effort,
again a simple rule is obtained. Finally, in Sect.~IV we discuss our
analysis.

\section{RC to polarized baryon semileptonic decays with
positively-charged lepton emission}

When $A$ is not polarized the RC with $l^{+}$ emission in groups (7) and
(8) are easily obtained from the final results of groups (5) and (6) when $%
l^{-}$ is emitted, using the rule of Ref.~\cite{c2}. However, this rule does
not apply to the RC to the part of the differential decay rate, and along
with it to the Dalitz plot (DP) containing the polarization of $A$. It is the
purpose of this section to obtain the corresponding rule. To do this
requires that we review the calculation of RC at intermediate steps and
trace the changes introduced by $l^{+}$ emission. The rule will allow us to
use the final expressions with $l^{-}$ emission to obtain directly the
final result with $l^{+}$ emission.

\subsection{Virtual RC}

We shall first discuss the decays $A^{0}\rightarrow B^{-}l^{+}\nu_l $ of
group (8). The calculation of its virtual RC follows the same steps of
the corresponding calculation of $A^{0}\rightarrow B^{+}l^{-}\overline{\nu}_l$ of
group (6). These corrections are split into a finite, calculable, and
model-independent part and into a model-dependent one, which can be
absorbed into the form factors of the uncorrected amplitude ${\sf M}_0$. This
last is indicated by putting primes on the form factors and ${\sf M}_0$. Thus, the decay
amplitude with virtual radiative corrections turns out to be
${\sf M}_V = {\sf M}_{0}^{\prime}+{\sf M}_{v}^{i}$, where ${\sf M}_{0}^{\prime}= (G_V/\sqrt{2})
H_{\lambda}^{\prime }L_{\lambda }^{\mp }$, with $H_{\lambda }^\prime= {\overline u}_{2}
\left( p_{2}\right) W_{\lambda }^{\prime }u_{1}\left(
p_{1}\right)$, $L_{\lambda }^{-}=\overline{u}_lO_{\lambda }v_{\nu}$,
$L_{\lambda }^{+}=\overline{u}_{\nu}O_{\lambda }v_l$, whereas ${\sf M}_{v}^{i}$,
after integrations over the virtual photon four-momentum, is given by
\begin{eqnarray}
{\sf M}_{v}^{i}=\frac{\alpha }{\pi }\frac{G_V}{\sqrt{2}}H_{\lambda }\left[
L_{\lambda }^{\mp }\phi \pm \left\{
\begin{array}{l}
{\overline u}_{l}{\not \! p}_{2} O_{\lambda }v_\nu \\
\overline{u}_{\nu} O_{\lambda }{\not \!p}_{2}v_l
\end{array}
\right\} \phi^\prime \right],
\end{eqnarray}
On the other hand, $W_{\lambda }^{\prime}$ is defined as
\begin{eqnarray}
W_{\lambda }^{\prime }=\gamma_{\lambda }\left( f_{1}^{\prime
}+g_{1}^{\prime }\gamma_{5}\right) +\sigma_{\lambda \alpha }\left(
q_{\alpha }/M_{1}\right) \left( f_{2}^{\prime }+g_{2}^{\prime }\gamma
_{5}\right) +\left( q_{\lambda }/M_{1}\right) \left( f_{3}^{\prime
}+g_{3}^\prime \gamma_{5}\right),
\end{eqnarray}
with $O_{\lambda }=\gamma_{\lambda }\left( 1+\gamma_{5}\right)$. Our
metric and $\gamma$-matrix convention are those of Ref.~\cite{c1}. In (11) the
upper (lower) sign refers to the upper (lower) sign of $A^{0}\rightarrow
B^{\pm }l^{\mp }\nu_l$. The spinors $u_{1}$ and $u_{2}$ belong to $A$ and
$B$, respectively.

Our interest here is in polarized decaying baryons $A$ along $s_1$,
so we shall concentrate on this part of the transition probability. In ${\sf M}_V$
one must replace $u_{1} \to \Sigma ({\not \! s}_{1}) u_{1}$, with $\Sigma ({\not \! s}_{1}) =
(1-\gamma_5 {\not \! s}_1)/2$, square ${\sf M}_V$ and sum over spins. Extracting the part
that contains $s_1$, we obtain
\begin{eqnarray}
\sum_s \left| {\sf M}_V^{(s)} \right|^{2} & = & \frac{1}{2} \sum_s \left| {\sf M}_{0}^{\prime \; (s)}
\right|^{2} + \frac{1}{2} c_{2}{\rm Tr} \left[ \left({\not \! p}_{2}+M_{2}\right)
W_{\lambda }\gamma_{5} {\not \! s}_{1}\left({\not \! p}_{1}+M_{1}\right) {\overline W}_{\mu }\right]
\nonumber \\
& & \times \left\{ \left( {\rm Re} \ \phi \right) {\rm Tr} \left[{\not \! \, l}
\gamma_{\lambda }{\not \! p}_{\nu }\gamma_{\mu }\left( 1\pm \gamma_{5}\right) \right]
 + \left( {\rm Re} \ \phi^\prime \right) {\rm Tr} \left[ {\not \! p}_{2}\gamma_{\lambda }
{\not \! p}_{\nu }\gamma_{\mu }\left( 1\pm \gamma
_{5}\right) \right] \right\}.
\end{eqnarray}
The explicit forms of $\phi$, $\phi^\prime$, and the constant $c_2$
are not relevant here. They can be found in Ref.~\cite{c7}. All we need to know is
that $c_2$ is real and only the real part of $\phi$ and $\phi^\prime$
appear in (13). $l$ and $p_\nu$ are the four-momenta of $l^{\mp}$
and the accompanying neutrino, respectively.

Equation (13) is a real quadratic function of the form factors $f_{i}^{\prime }$
and $g_{i}^{\prime}$. If we assume momentarily $g_{i}^{\prime }=0$ we
obtain a hadronic trace containing only one $\gamma_{5}$. If instead we
assume $f_{i}^{\prime }=0$, we obtain $(\gamma_{5})^{3}=\gamma_{5}$ in this
trace. Thus, the hadronic part of (13) containing non-interference $f_i^\prime f_j^\prime$
and $g_i^\prime g_j^\prime$ products is imaginary. The interference products
$f_i^\prime g_j^\prime$ give $(\gamma_5)^2=1$ in this trace. Accordingly, the
part of the trace containing these products is real.

The leptonic trace also contains a real part and an imaginary one, this
latter coming from the $\gamma_{5}$ contribution. Since Eq.~(13) is
necessarily real and the double sign is attached to the $\gamma_{5}$ in the
leptonic trace, we can now obtain the rule we are looking for: {\it to use the
results of the polarization part of $l^{-}$ emission of decays (6) for
$l^{+}$ emission of decays (8), one must reverse the signs of all the
non-interference products of form factors and keep the same signs in the
interference products}. This rule should be contrasted with the rule in the
unpolarized decay rate~\cite{c2}; it is the opposite, so to speak. We did not
discuss the changes in the ${\sf M}_{0}^{\prime \, (s)}$ contribution to
Eq.~(13). One can readily see that one obtains the same rule. Also, the rule
to connect decays (7) with decays (5) is immediate.

\subsection{Bremsstrahlung RC}

Again we shall discuss first the group of decays $A^{0}\rightarrow B^{\pm
}l^{\mp }\nu_l \gamma$, where $\gamma$ is a real photon of
four-momentum $k$. The transition amplitude contains three terms,
\begin{eqnarray}
{\sf M}_{B}={\sf M}_{B_{1}}+{\sf M}_{B_{2}}+{\sf M}_{B_{3}}. \nonumber
\end{eqnarray}

The Low theorem~\cite{c6}, in the form presented by Chew~\cite{c8}, allows us to get
\begin{eqnarray}
{\sf M}_{B_{1}} &=&\pm \frac{eG_V}{\sqrt{2}}\epsilon_{\mu }\left(
\frac{l_{\mu }}{l \cdot k}-\frac{p_{2\mu }}{p_{2}\cdot k}\right)
H_{\lambda }L_{\lambda }^{\mp }, \\
{\sf M}_{B_{2}} &=&\pm \frac{eG_V}{\sqrt{2}}\epsilon_{\mu }H_{\lambda
}\frac{1}{2l \cdot k}\left\{
\begin{array}{l}
\overline{u}_l\gamma_{\mu }{\not \! k}O_{\lambda }v_{\nu } \\
\overline{u}_{\nu }O_{\lambda }{\not \! k}\gamma_{\mu }v_l
\end{array}
\right\}, \\
{\sf M}_{B_{3}} &=&\pm \frac{eG_V}{\sqrt{2}}\epsilon_{\mu } {\overline u}_{2}
T_{\mu \lambda }u_{1}L_{\lambda }^{\mp }.
\end{eqnarray}
Here $e$ (a negative number) is the charge of $l^{-}$ and $\epsilon_\mu$ is the
polarization four-vector of $\gamma$. $H_{\lambda}$, $L_\lambda^\mp$, and the upper
and lower signs have the same meaning as before. The tensor $T_{\mu \lambda}$ is given by
\begin{eqnarray}
T_{\mu \lambda} & = & \frac{1}{2p_2 \cdot k} \left[ -\gamma_\mu {\not \! k} -
\frac{\kappa_{2}}{e_{2}}\sigma_{\mu \alpha }k_{\alpha }\left(
{\not \! p}_{2}+{\sf M}_{2}\right) \right] W_{\lambda } \nonumber \\
&  & \mbox{} + \frac{1}{2p_{1}\cdot k}W_{\lambda }\frac{\kappa_{1}}{e_{2}}\left(
{\not \! p}_{1}+M_{1}\right) \sigma_{\mu \alpha }k_{\alpha } \nonumber \\
& & \mbox{} + \left( \frac{p_{2\mu }k_{\rho }}{p_{2}\cdot k}-g_{\mu \rho }\right) \left(
\sigma_{\lambda \rho }\frac{f_{2}+g_{2}\gamma_{5}}{M_{1}}+g_{\lambda \rho }%
\frac{f_{3}+g_{3}\gamma_{5}}{M_{1}}\right),
\end{eqnarray}
where $\kappa_{1}$ and $\kappa_{2}$ are the anomalous magnetic moments of $A^{0}$
and $B^{\pm }$, respectively, and $e_{2}$ is the charge of $B^{\pm }$.
Again we shall concentrate on the part of the bremsstrahlung transition rate
that contains the polarization $s_1$ of $A^{0}$. Introducing $\Sigma ({\not \! s}_{1}) u_1$
in ${\sf M}_B$, squaring ${\sf M}_{B}$, and summing over
spins, one can extract the polarization part of the transition probability.
The result is
\begin{eqnarray}
\sum_{s,\epsilon} \left| {\sf M}_{B}^{(s)} \right|^{2} & = & c_{3} {\rm Re} \sum_{\epsilon}
\epsilon_{\alpha } \epsilon_{\beta} \left\{ {\rm Tr} \left[ \left({\not \! p}_2 + M_2 \right)
W_{\lambda }\gamma_{5} {\not \! s}_{1}\left({\not \! p}_1 + M_1 \right)
\overline{W}_{\mu }\right] \right. \nonumber \\
&  & \mbox{} \times \left[ I^{\alpha }I^{\beta } {\rm Tr} \left[ {\not \! \, l} \gamma_\lambda
{\not \! p}_{\nu} \gamma_{\mu }\left( 1\pm \gamma_{5}\right) \right] \nonumber  \right.\\
&  & \mbox{} + \left( \frac{1}{2l \cdot k}\right)^{2} {\rm Tr} \left[ {\not \! k}\gamma
_{\beta }{\not \! \, l}\gamma_{\alpha }{\not \! k}\gamma_{\lambda }{\not \! p}_{\nu
}\gamma_{\mu }\left( 1\pm \gamma_{5}\right) \right] \nonumber \\
& & \mbox{} + \left. \frac{I^\alpha}{l \cdot k} {\rm Tr} \left[ {\not \! k}\gamma_{\beta }
{\not \! \, l}\gamma_{\lambda }{\not \! p}_{\nu }\gamma_{\mu }( 1\pm \gamma
_{5}) \right] \right] \nonumber \\
&  & \mbox{} + {\rm Tr} \left[ ({\not \! p}_{2}+M_{2}) T_{\alpha \lambda }\gamma_{5}%
{\not \! s}_{1}\left( {\not \! p}_{1}+M_{1}\right) \overline{W}_{\mu }\right] \nonumber \\
&  & \mbox{} \left. \times \left[ 2I_{\beta } {\rm Tr} \left[{\not \! \, l}\gamma_{\lambda }%
{\not \! p}_{\nu }\gamma_{\mu }\left( 1\pm \gamma_{5}\right) \right] +\frac{1}{%
l \cdot k} {\rm Tr} \left[{\not \! k}\gamma_{\beta }{\not \! \, l}\gamma_{\lambda}
{\not \! p}_{\nu }\gamma_{\mu }\left( 1\pm \gamma_{5}\right) \right] \right] \right\}.
\end{eqnarray}
Here $c_3$ is an overall real constant containing $\alpha$, $G_V$, etc.,
and $I_{\alpha}=( l_{\alpha }/l \cdot k) - p_{2_\alpha}/p_{2}\cdot k)$. The first three terms in
Eq.~(18) come from the squares and interference of ${\sf M}_{B_1}$ and ${\sf M}_{B_2}$.
The hadronic trace in them is the same one as in Eq.~(13). Therefore, the
non-interference products $f_{i}^{\prime }f_{j}^{\prime }$ and $%
g_{i}^{\prime }g_{j}^{\prime }$ are accompanied by a $\gamma_{5}$ and the
interference products $f_{i}^{\prime } g_{j}^{\prime }$ are not. The
reality condition on Eq.~(18) and the double sign in front of the $\gamma_5$ in the leptonic
trace lead to the same rule of the virtual RC.

The last two terms of Eq.~(18) contain T$_{\alpha \lambda }$ in the hadronic
trace. Using Eq.~(17) they can be rearranged into the sum of
\begin{eqnarray}
&  & {\rm Tr} \left[( {\not \! p}_2 + M_2) \gamma_\alpha {\not \! k} W_\lambda
\gamma_{5} { \not \! s}_1 ( {\not \! p}_1 + M_1) \overline{W}_\mu \right], \nonumber \\
&  & {\rm Tr} \left[( {\not \! p}_2 + M_2) \sigma_{\alpha \beta} k_\rho ( {\not \!p}_2
+ M_2) W_\lambda \gamma_5 {\not \! s}_1({\not \!p}_1 + M_1) \overline{W}_\mu \right], \nonumber \\
&  & {\rm Tr} \left[( {\not \! p}_2 + M_2)) \sigma_{\lambda \alpha}(f_2+g_2 \gamma_5)
\gamma_5 {\not \! s}_1 ( {\not \! p}_1 + M_1) \overline{W}_\mu \right], \nonumber
\end{eqnarray}
and
\begin{eqnarray}
{\rm Tr} \left[( {\not \! p}_2 + M_2) g_{\lambda \alpha} (f_3+g_3 \gamma_5) \gamma_5 {\not \! s}_1
({\not \! p}_1 + M_1) \overline{W}_\mu \right].
\end{eqnarray}

Let us now follow the same steps as before. Assuming momentarily all $g_i=0$ or all $f_i=0$
we see that these traces contain either $\gamma_5$ or $(\gamma_5)^3$. So the traces 
with non-interference products are imaginary numbers. The traces with interference products
do not contain $\gamma_5$ and are accordingly real numbers. Then, the reality condition
on Eq.~(18) and the position of the double sign in front of the $\gamma_{5}$
in the leptonic trace lead to the same rule as before.

Collecting all the previous results we can establish the complete rule to obtain the RC to
the polarization part of decay $A^{0}\rightarrow B^{-}l^+ \nu_l$ directly from the final
RC to the polarization part of $A^{0}\rightarrow B^{+}l^{-}\overline{\nu}_l$: {\it one must
reverse the signs of the non-interference products $f_{i}^{\prime }f_{j}^{\prime }$ and 
$g_i^\prime g_j^\prime$ and keep the same sign of the interference
products $f_i^\prime g_j^\prime$}. It is clear that this rule
covers contributions of orders $\left( \alpha /\pi \right) \left(
q/M_{1}\right)^{0}$ and $\left( \alpha /\pi \right) \left( q/M_{1}\right)$.

The same analysis applies to the decays $A^{\pm }\rightarrow B^{0}l^{\pm
}\nu_l $ of groups (5) and (7) and one comes to the same rule: {\it the RC to the
polarization part of $A^{+}\rightarrow B^{0}l^{+}\nu_l $ are obtained from
the final result of the RC to the polarization part of $A^- \rightarrow B^0l^-\overline{\nu}_l$
by reversing the signs in front of the products $f_i^\prime f_j^\prime$ and $g_i^\prime
g_j^\prime$ and keeping the same sign in front of the products $f_i^\prime g_j^\prime$}.

As already mentioned, this rule is, so to speak, the opposite to the rule
that applies in the unpolarized decay rate. In this case the interference
products of form factors must reverse their signs, while the
non-interference ones preserve their signs. Equivalently, one can cover both
polarized and unpolarized cases by restating the rules as {\it to change the
signs of all $g_{i}$ form factors and the sign of $s_1$}. In the literature there
exists another rule~\cite{c9} to change the results with $l^-$ emission into the final
results with $l^+$ emission. This rule is given in terms of the lepton four-momenta. One
clearly sees that it is not of practical use when RC are incorporated.

To close this section, let us remark that the rule obtained is applicable to any $l^+$
$(e^{+},\mu^{+},\tau^{+})$. It applies in any Lorentz frame, and it is valid in both
the three-body and four-body regions of the DP.

\section{RC to semileptonic decays with double-charge baryons}

To study the RC to decays in groups (9) and (10) requires that we start at
the graph level, extending the work of Refs.~\cite{c5,c6}. For
definiteness, we shall discuss decays of type (9) and at the end we shall
include the decays of type (10).

\subsection{Virtual RC}

The Feynman diagrams for decays (9) are displayed in Fig.~1. The blobs stand for
the effects of strong interactions and details of weak interactions. Our notation
and conventions are those of Ref.~\cite{c5}.

\begin{figure}
\centerline{\epsfxsize = 14.8cm \epsfbox{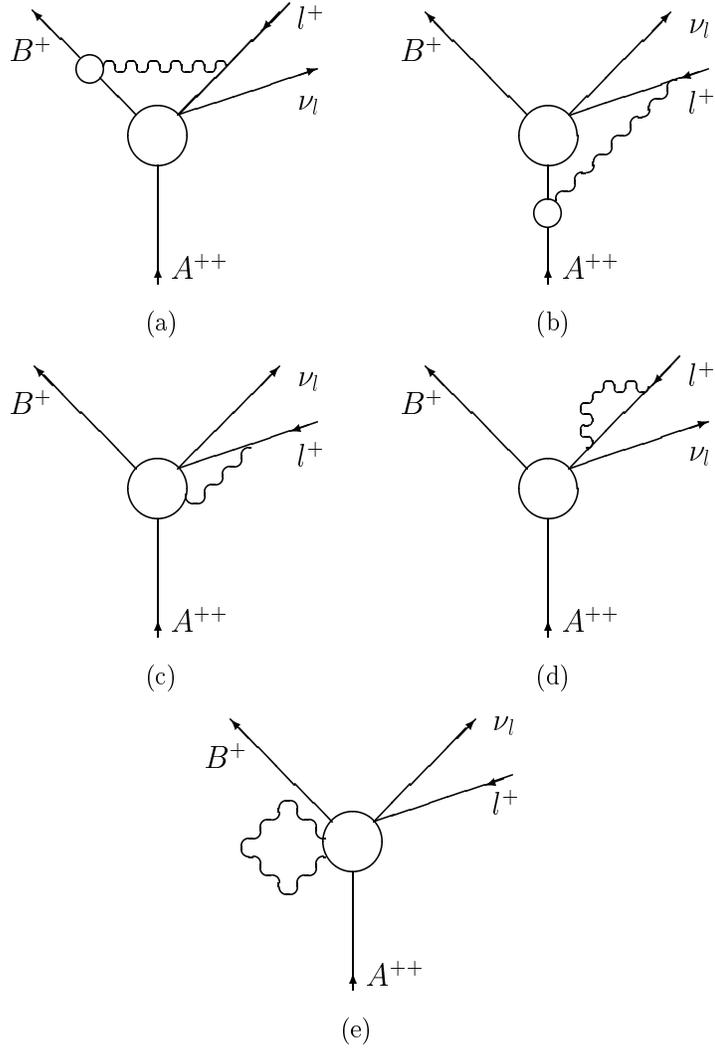}}
\caption{Feynman graphs for virtual RC to BSD. The blobs stand for
strong interaction effects and details of weak interactions.}
\end{figure}

The virtual RC can be split into a finite and model-independent part and
into a model-dependent one. In diagrams 1(a), 1(b), and 1(c) the virtual photon
is interchanged between $l^{+}$ and a charge line within $B^{+}$, within $A^{++}$,
and within the weak vertex, respectively. The three diagrams
lead to a transition amplitude that can be cast into the form
\begin{eqnarray}
{\sf M}_{v_{1}} & = & -\frac{G_V}{\sqrt 2} \frac{\alpha}{4\pi^3} \frac{1}{i}
\int d^{4}k D_{\mu \alpha}(k) \left\{ \overline{u}_{2} \left[ 2W_{\lambda }(p_2,p_1)
\frac{(2p_{1\mu}-k_\mu)}{k^2-2p_1 \cdot k+i\varepsilon} + T_{\mu \lambda}^+
(p_1,p_2,k) \right. \right. \nonumber \\
&  & \mbox{} + \left. \left. \left( \frac{2p_{2\mu }+k_{\mu }}{k^{2}+2p_{2}\cdot k+i\varepsilon}
W_\lambda (p_2,p_1) + T_{\mu \lambda}^0 (p_1,p_2,k) \right) \right] u_1 \right\}
\overline{u}_\nu O_\lambda \frac{2l_\alpha - {\not \! k} \gamma_\alpha}
{k^2-2l \cdot k+i\varepsilon}v_l.
\end{eqnarray}
Here $l$, $k$, $p_2$, and $p_1$ are the four-momenta of $l^+$, the virtual photon, $B^+$,
and $A^{++}$, respectively. All the model dependence in these diagrams is contained in the
tensors $T_{\mu \nu}^+$ and $T_{\mu \nu}^0$. Their upper indices will be explained shortly. The
other terms are model-independent.

Diagram 1(d), after wavefunction renormalization, leads to the amplitude
\begin{eqnarray}
{\sf M}_{v_{2}} = \frac{\alpha }{8\pi^{3}i}\frac{G_V}{\sqrt{2}}
\overline{u}_{2}W_{\lambda }( p_{1},p_{2}) u_1 \int d^4 k D_{\mu\alpha}( k)
\overline{u}_\nu O_\lambda \left({\not \! \, l}- m\right)
\frac{(2l_\mu+\gamma_\mu{\not \! k}){\not \! \, l}(2l_\alpha+{\not \! k}\gamma_\alpha)}
{2m^{2}\left( k^{2}+2l \cdot k+i\varepsilon \right)^{2}}_{{}}v_l.
\end{eqnarray}
Notice that here the RC is contained only within the lepton covariant.

The last diagram 1(e) contains~\cite{c10} a convection-convection contribution
${\sf M}_{v_3}^{c}$ which is model-independent and implements gauge invariance when
it is added to Eqs.~(20) and (21). The complete amplitude of Fig.~1(e) is then
\begin{eqnarray}
{\sf M}_{v_3}={\sf M}_{v_3}^{c}+{\sf M}_{v_3}^{\prime },
\end{eqnarray}
where
\begin{eqnarray}
{\sf M}_{v_{3}}^{c} & = & \frac{G_V}{\sqrt{2}}\frac{\alpha }{8\pi^{3}i}\overline{u}_2
W_{\lambda }(p_2,p_1) u_1\overline{u}_\nu O_\lambda v_l\int d^4kD_{\mu \alpha}(k) \nonumber \\
&  & \mbox{}\times \left\{ -\frac{( 2p_{2}-k)_\mu \left( 2p_{2}-k\right)
_{\alpha }}{\left( k^{2}-2p_{2}\cdot k+i\varepsilon \right)^{2}}+2\frac{%
\left( 2p_{1}-k\right)_{\mu}\left( 2p_{1}-k\right)_{\alpha }}{\left(
k^{2}-2p_{1}\cdot k+i\varepsilon \right)^{2}}\right\}.
\end{eqnarray}
All the model dependence of this diagram is contained in ${\sf M}_{v3}^{\prime }$.

The analysis of Ref.~\cite{c5} to deal with the model-dependent parts contained in
Eqs.~(20) and (22), limited to small $q$, can be extended as shown in Ref.~\cite{c11} to include
contributions of order $(\alpha/\pi)(q/M_1)$. The result of this extension is that up to this order all
the model dependence has the same form as the uncorrected ${\sf M}_0$ and this
allows that it be completely absorbed into the six already existing form
factors. We indicate this by putting a prime on each form factors and on ${\sf M}_{0}$, too.

Let us deal with the model-independent parts in Eqs.~(20), (21), and (23).
The way Eq.~(20) is written makes it easy to see that it can be rearranged
into
\begin{eqnarray}
{\sf M}_{v_{1}}( A^{++}\rightarrow B^{+}l^{+}\nu_l)
=2{\sf M}_{v_{1}}( A^{+}\rightarrow B^{0}l^{+}\nu_l)
-{\sf M}_{v_{1}}( A^{0}\rightarrow B^{-}l^{+}\nu_l).
\end{eqnarray}

This amplitude has been rewritten as a linear combination of the
amplitudes of the two BSD indicated in parentheses on the rhs. This
explains the choice of upper indices on the $T_{\mu \nu}$ of Eq.~(20).
Notice the factor two and the minus sign in this rhs. The Coulomb
contribution, attractive in $A^{0}\rightarrow B^{-}l^{+}\nu_l$, becomes
repulsive in $A^{++}\rightarrow B^{+}l^{+}\nu_l$, as expected.

Equation (21) can be rearranged analogously. The amplitude ${\sf M}_{v_{2}}$ here is the
same as ${\sf M}_{v_{2}}$ of $A^{+}\rightarrow B^{0}l^{+}\nu_l$ and of
$A^{0}\rightarrow B^{-}l^{+}\nu_l$. So, one immediately gets
\begin{eqnarray}
{\sf M}_{v_{2}}( A^{++}\rightarrow B^{+}l^{+}\nu_l)
=2{\sf M}_{v_{2}}( A^{+}\rightarrow B^{0}l^{+}\nu_l)
-{\sf M}_{v_{2}}( A^{0}\rightarrow B^{-}l^{+}\nu_l).
\end{eqnarray}

Equation (23) can also be cast into the same linear combination, namely,
\begin{eqnarray}
{\sf M}_{v_{3}}( A^{++}\rightarrow B^{+}l^{+}\nu_l)
=2{\sf M}_{v_{3}}( A^{+}\rightarrow B^{0}l^{+}\nu_l)
-{\sf M}_{v_{3}}( A^{0}\rightarrow B^{-}l^{+}\nu_l).
\end{eqnarray}

The transition amplitude with virtual RC becomes,
\begin{eqnarray}
{\sf M}_{V}( A^{++}\rightarrow B^{+}l^{+}\nu_l) =2\left[
{\sf M}_{0}^{\prime }+{\sf M}_{v}( A^{+}\rightarrow B^{0}l^{+}\nu_l)
\right] -\left[ {\sf M}_{0}^{\prime }+{\sf M}_{v}( A^{0}\rightarrow B^{-}l
^{+}\nu_l) \right],
\end{eqnarray}
where we added and subtracted ${\sf M}_{0}^{\prime }$, and ${\sf M}_{v}$ stand for the
sum of the three virtual model-independent RC. The square brackets
contain the transition amplitudes with virtual RC of $A^{+}\rightarrow
B^{0}l^{+}\nu_l$ and $A^{0}\rightarrow B^{-}l^{+}\nu_l$. Equation~(27) can be compactly rewritten as
\begin{eqnarray}
{\sf M}_{V}( A^{++}\rightarrow B^{+}l^{+}\nu_l) =2{\sf M}_{V}(A^{+}\rightarrow B^{0}l^{+}
\nu_l) -{\sf M}_{V} ( A^{0}\rightarrow B^{-}l^{+}\nu_l).
\end{eqnarray}

This analysis can be repeated step by step for $A^{+}\rightarrow B^{++}l^{-} \overline{\nu}_l$. 
The result is
\begin{eqnarray}
{\sf M}_{V}( A^{+}\rightarrow B^{++}l^{-}\overline{\nu}_l ) =2{\sf M}_{V} (
A^{0}\rightarrow B^{+}l^{-}\overline{\nu}_l) -{\sf M}_{V} ( A^{-}\rightarrow
B^{0}l^{-} \overline{\nu}_l).
\end{eqnarray}
Now the Coulomb interaction is attractive and the double charge of $B^{++}$
is taken care of by the factor two in front of the first term on the rhs of
Eq.~(29).

All the integrals over the virtual photon four-momentum required to get the
virtual RC to $A^{++}\rightarrow B^{+}l^{+}\nu_l$ and $A^{+}\rightarrow
B^{++}l^{-}\overline{\nu}_l$ can be taken from previous work \cite{c2,c7,c12}. ${\sf M}_{V}(
A^{+}\rightarrow B^{0}l^{+}\nu_l)$ and ${\sf M}_{V}( A^{0}\rightarrow B^{-}l^{+}\nu_l)$
are given in Ref.~\cite{c2}. ${\sf M}_{V}(A^{0}\rightarrow B^{+}l^{-}\overline{\nu}_l)$ and
${\sf M}_{V}( A^{-}\rightarrow B^{0}l^{-}\overline{\nu}_l)$ are given in Refs.~\cite{c7,c12}.

To obtain the differential decay rates and the DP corresponding to
amplitudes (28) and (29), one follows the usual steps of squaring, summing
and averaging over spins, and so on. For polarized decaying baryons one must
replace $u_{1}$ by $\Sigma ({\not \! s}_{1}) u_1$. However, one must
remember that the masses and form factors to be used in the rhs of
Eqs.~(28) and (29) are those of their lhs. We shall now show that at the
differential decay rate level one obtains to first order in $\alpha/\pi$ the same linear
combinations as at the amplitude level. Let us discuss again $A^{++}\rightarrow B^{+}l^{+}\nu_l$.

From Ref.~\cite{c2} one has that
\begin{eqnarray}
{\sf M}_{V}( A^{0}\rightarrow B^{-}l^{+}\nu_l) ={\sf M}_{0}^{\prime}+
\frac{\alpha}{\pi}  ({\sf M}_{0}\phi +{\sf M}_{p_{2}}\phi^{\prime })
\end{eqnarray}
and
\begin{eqnarray}
{\sf M}_{V}( A^{+} \rightarrow B^{0}l^{+}\nu_l) = {\sf M}_{0}^{\prime}+
\frac{\alpha }{2\pi }( {\sf M}_{0}\Phi +{\sf M}_{p_{1}}\Phi^{\prime }).
\end{eqnarray}

Equation (28) leads to
\begin{eqnarray}
\sum \left| {\sf M}_{V}( A^{++} \rightarrow B^{+}l^{+}\nu_l) \right|^2
& = & 4 \sum \left| {\sf M}_{V}( A^{+}\rightarrow B^{0}l^{+}\nu_l) \right|^{2}
+ \sum \left| {\sf M}_{V}( A^{0}\rightarrow B^{-}l
^{+}\nu_l) \right|^{2} \nonumber \\
&  & \mbox{} - 4{\rm Re} \sum {\sf M}_{V}( A^{+}\rightarrow B^{0}l^{+}\nu_l)
\overline{{\sf M}}_{V}( A^{0}\rightarrow B^{-}l^{+}\nu_l).
\end{eqnarray}

The bar means transpose conjugate. The cross term becomes, after
substituting (30) and (31) and keeping only first order terms in $(\alpha/\pi)$,
\begin{eqnarray}
& & -4 {\rm Re}\sum {\sf M}_{V}\left( A^{+}\rightarrow B^{0}l^{+}\nu_l \right)
\overline{{\sf M}}_{V}\left( A^{0}\rightarrow B^{-}l^{+}\nu_l \right) \nonumber \\
& & = -2\left\{ \left[ \sum \left| {\sf M}_{0}^{\prime }\right|^{2}+\frac{\alpha}{2\pi}
2 {\rm Re} \sum {\sf M}_{0}^{\prime }\overline{({\sf M}_{0}\Phi +{\sf M}_{p_{1}}\Phi^{\prime }) }
\right] \right. \nonumber \\
& & \mbox{\hglue1.0truecm} + \left. \left[ \sum \left| {\sf M}_{0}^{\prime } \right|^{2}
+ \frac{\alpha}{\pi} 2 {\rm Re} \sum ( {\sf M}_{0}\phi +{\sf M}_{p_{2}}\phi^{\prime})
\overline{{\sf M}_{0}^\prime} \right] \right\}.
\end{eqnarray}
One can recognize within the brackets on the rhs the squares of the
amplitudes (30) and (31) to first order in $(\alpha/\pi)$. Thus Eq.~(33) becomes
\begin{eqnarray}
&  & -4 {\rm Re}\sum {\sf M}_{V}\left( A^{+}\rightarrow B^{0}l^{+}\nu_l \right)
\overline{{\sf M}}_{V}(A^{0}\rightarrow B^{-}l^{+}\nu_l) \nonumber \\
& & \mbox{\hglue1.0truecm} = - 2 \sum \left| {\sf M}_{V}( A^{+}\rightarrow B^{0}l^{+}\nu_l)
\right|^{2}-2\sum \left| {\sf M}_{V}( A^{0}\rightarrow B^{-}l^{+}\nu_l) \right|^{2}.
\end{eqnarray}

Collecting results the differential decay rate for $A^{++}\rightarrow
B^{+}l^{+}\nu_l$ becomes
\begin{eqnarray}
d\Gamma_{V}(A^{++}\rightarrow B^{+}l^{+}\nu_l) =2d\Gamma
_{V} (A^{+}\rightarrow B^{0}l^{+}\nu_l) -d\Gamma_{V}(A^{0}\rightarrow B^{-}l^{+}\nu_l).
\end{eqnarray}

Exactly the same steps and using again Ref.~\cite{c2} lead to
\begin{eqnarray}
d\Gamma_{V} (A^{+}\rightarrow B^{++}l^{-} \overline{\nu}_l) =2d\Gamma
_{V} (A^{0}\rightarrow B^{+}l^{-}\overline{\nu}_l) -d\Gamma_{V}(A^{-}\rightarrow 
B^{0}l^{-}\overline{\nu}_l).
\end{eqnarray}

Equations (35) and (36) show that we can use directly the decay rates previously
obtained to get the decay rates of groups (9) and (10), without having to
repeat the calculation of neither the virtual photon integrals nor the
traces. Of course the form factors and masses appropriate to these decays
must be used in the previous results.

\subsection{Bremsstrahlung RC}

\begin{figure}
\centerline{\epsfxsize = 14.8cm \epsfbox{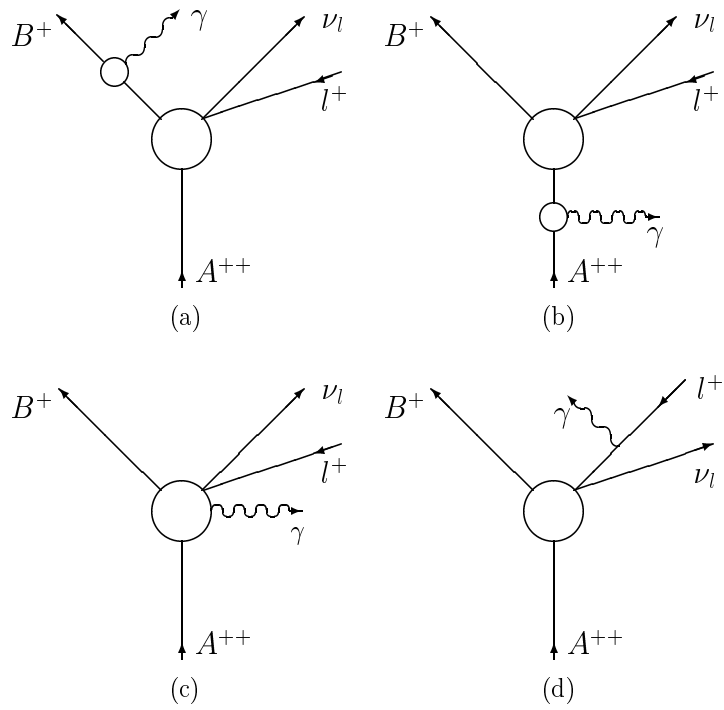}}
\caption{Feynman graphs for BSD with real photon emission. The blobs stand for
strong interaction effects and details of weak interactions.
}
\end{figure}


The Feynman diagrams with real photon emission of decays of group (9) are displayed in
Fig.~2. The blob stands for strong-interaction effects and details of weak interactions.
The Low theorem~\cite{c6} allows the inclusion of terms of up to order $(\alpha/\pi)(q/M_1)$
in a model-independent fashion. We shall use the approach of Chew~\cite{c8} to use this
theorem. The transition amplitude of the diagrams in Fig.~2 can be split into three contributions,
\begin{eqnarray}
{\sf M}_{B}( A^{++}\rightarrow B^{+}l^{+}\nu_l \gamma) = {\sf M}_{B_1}+{\sf M}_{B_2}+{\sf M}_{B_3},
\end{eqnarray}
with
\begin{eqnarray}
{\sf M}_{B_{1}} & = & -e{\sf M}_{0}\left[ \frac{\epsilon \cdot l }{l \cdot k}-2\frac{%
\epsilon \cdot p_{1}}{p_{1}\cdot k}+\frac{\epsilon \cdot p_{2}}{p_{2}\cdot k}\right], \\
{\sf M}_{B_{2}} & = & -e\frac{G_V}{\sqrt{2}}\epsilon_{\mu }\,\overline{u}_{2}
W_{\lambda } u_{1}\,\overline{u}_{\nu}O_{\lambda }\frac{\not{k}%
\gamma_{\mu}}{2l \cdot k} v_l,
\end{eqnarray}
and
\begin{eqnarray}
{\sf M}_{B_{3}} & = & \frac{G_V}{\sqrt{2}}\overline{u}_{\nu} O_{\lambda }\,v_l\,\epsilon_{\mu }
\,\overline{u}_{2}\,\left\{
-2eW_{\lambda }\frac{{\not \! k}\gamma_{\mu }}{2p_{1}\cdot k}-\kappa
_{1}W_{\lambda }\frac{{\not \! p}_{1}+M_{1}}{2p_{1}\cdot k}\sigma_{\mu \nu
}k_{\nu }\right. \nonumber \\
&  & \mbox{} -e\frac{\gamma_{\mu } {\not \! k}}{2p_{2}\cdot k}W_{\lambda }+\kappa_{2}\sigma
_{\mu \nu }k_{\nu }\frac{{\not \! p}_{2}+M_{2}}{2p_{2}\cdot k}W_{\lambda } \nonumber \\
&  & \mbox{} -2e\left[ \frac{p_{1\mu }\,}{p_{1}\cdot k}k_{\rho }-g_{\mu \rho }\right] %
\left[ \sigma_{\rho \lambda }\frac{f_{2}+g_{2}\gamma_{5}}{M_{1}}+g_{\rho
\lambda }\frac{f_{3}+g_{3}\gamma_{5}}{M_{1}}\right] \nonumber \\
&  & \mbox{} + \left. e\left[ \frac{p_{2\mu }\,}{p_{2}\cdot k}k_{\rho }-g_{\mu \rho }%
\right] \left[ \sigma_{\rho \lambda }\frac{f_{2}+g_{2}\gamma_{5}}{M_{1}}%
+g_{\rho \lambda }\frac{f_{3}+g_{3}\gamma_{5}}{M_{1}}\right] \right\} u_{1}.
\end{eqnarray}
Here $\epsilon_{\mu}$ is the photon polarization, $e$ is the charge of
the electron ($e<0$), $\kappa_{1}$ and $\kappa_{2}$ are the
anomalous magnetic moments of $A$ and $B$, respectively, and $f_{i}$, $g_{i}$
are form factors.

By adding and subtracting appropriate terms, we can rearrange these equations
into
\begin{eqnarray}
{\sf M}_{B_{1}}=-2e{\sf M}_{0}\left[ \frac{\epsilon \cdot l }{l \cdot k}-\frac{%
\epsilon \cdot p_{1}}{p_{1}\cdot k}\right] -e{\sf M}_{0}\left[ \frac{%
\epsilon \cdot p_{2}}{p_{2}\cdot k}-\frac{\epsilon \cdot l }{l
\cdot k}\right],
\end{eqnarray}
\begin{eqnarray}
{\sf M}_{B_{2}}=-2e\frac{G_V}{\sqrt{2}}\epsilon_{\mu }\,\overline{u}%
_{2}\,W_{\lambda }\,u_{1}\,\overline{u}_{\nu}O_{\lambda }\frac{{\not \!k}
\gamma_{\mu}}{2l \cdot k} v_l-\left( -e\right) \frac{G_V}{%
\sqrt{2}}\epsilon_{\mu }\,\overline{u}_{2}\,W_{\lambda }\,u_{1}\,%
\overline{u}_{\nu}O_{\lambda }\frac{{\not \! k}\gamma_{\mu }}{2l \cdot k} v_l,
\end{eqnarray}
\begin{eqnarray}
{\sf M}_{B_3} & = & \frac{G_V}{\sqrt 2} \epsilon_\mu \overline{u}_2 \left\{ -2eW_\lambda
\frac{{\not \! k}\gamma_{\mu }}{2p_1 \cdot k}-2\kappa_1 W_\lambda
\frac{{\not \! p}_{1}+M_{1}}{2p_{1}\cdot k} \sigma_{\mu \nu }k_{\nu } + 2\kappa_{2}
\sigma_{\mu \nu }k_{\nu }\frac{{\not \! p}_{2}+M_{2}}{2p_{2}\cdot k}
W_{\lambda } \right. \nonumber \\
&  & \mbox{} -2e\left[ \frac{p_{1\mu }\,}{p_{1}\cdot k}k_{\rho }-g_{\mu \rho} \right]
\left[ \sigma_{\rho \lambda }\frac{f_{2}+g_{2}\gamma_{5}}{M_{1}}%
+g_{\rho \lambda }\frac{f_{3}+g_{3}\gamma_{5}}{M_{1}}\right] \nonumber \\
&  & \mbox{} -e\frac{\gamma_{\mu }{\not \!k}}{2p_{2}\cdot k}W_{\lambda }+\kappa
_{1}W_{\lambda }\frac{{\not \! p}_{1}+M_{1}}{2p_{1}\cdot k}\sigma_{\mu \nu
}k_{\nu }-\kappa_{2}\sigma_{\mu \nu }k_{\nu }\frac{{\not \! p}_{2}+M_{2}}{%
2p_{2}\cdot k}W_{\lambda } \nonumber \\
&  & \mbox{} +\left. e\left[ \frac{p_{2\mu }\,}{p_{2}\cdot k}k_{\rho }-g_{\mu \rho }%
\right] \left[ \sigma_{\rho \lambda }\frac{f_{2}+g_{2}\gamma_{5}}{M_{1}}%
+g_{\rho \lambda }\frac{f_{3}+g_{3}\gamma_{5}}{M_{1}}\right] \right\} u_{1}%
\overline{u}_{\nu }\,O_{\lambda }\,v_l.
\end{eqnarray}

It is now easy to identify the bremsstrahlung amplitudes of decays of groups
(7) and (8). Equations (41)-(43) can be expressed as
\begin{eqnarray}
{\sf M}_{B_1}( A^{++}\rightarrow B^{+}l^{+}\nu_l \gamma )
=2{\sf M}_{B_1}( A^{+}\rightarrow B^{0}l^{+}\nu_l \gamma )
-{\sf M}_{B_1} ( A^{0}\rightarrow B^{-}l^{+}\nu_l \gamma ),
\end{eqnarray}
\begin{eqnarray}
{\sf M}_{B_2}( A^{++}\rightarrow B^{+}l^{+}\nu_l \gamma )
=2{\sf M}_{B_2}( A^{+}\rightarrow B^{0}l^{+}\nu_l \gamma )
-{\sf M}_{B_2}( A^{0}\rightarrow B^{-}l^{+}\nu_l \gamma ),
\end{eqnarray}
\begin{eqnarray}
{\sf M}_{B_3}( A^{++}\rightarrow B^{+}l^{+}\nu_l \gamma )
=2{\sf M}_{B_3}( A^{+}\rightarrow B^{0}l^{+}\nu_l \gamma )
-{\sf M}_{B_3}( A^{0}\rightarrow B^{-}l^{+}\nu_l \gamma ).
\end{eqnarray}

Collecting terms, the amplitude ${\sf M}_{B}$ of Eq.~(37) becomes
\begin{eqnarray}
{\sf M}_{B}( A^{++}\rightarrow B^{+}l^{+}\nu_l \gamma ) =2{\sf M}_{B} (
A^{+}\rightarrow B^{0}l^{+}\nu_l \gamma ) -{\sf M}_{B}(A^{0}\rightarrow B^{-}l^{+}\nu_l \gamma ).
\end{eqnarray}
Again, like in the virtual RC we have expressed the amplitude of the process
$A^{++}\rightarrow B^{+}l^{+}\nu_l \gamma$ as a linear combination of the
amplitudes of the processes $A^{+}\rightarrow B^{0}l^{+}\nu_l \gamma$ and
$A^{0}\rightarrow B^{-}l^{+}\nu_l \gamma$. These latter can be found in
Ref.~\cite{c2}, with a minor change of notation.

In the same way we can study the bremsstrahlung amplitude of decays $%
A^{+}\rightarrow B^{++}l^{-}\overline{\nu}_l $ of group (10). The result is
\begin{eqnarray}
{\sf M}_{B}( A^{+}\rightarrow B^{++}l^{-}\overline{\nu}_l \gamma ) =2{\sf M}_{B}(
A^{0}\rightarrow B^{+}l^{-}\overline{\nu}_l \gamma) -{\sf M}_{B}(
A^{-}\rightarrow B^{0}l^{-}\overline{\nu}_l \gamma ).
\end{eqnarray}
The ${\sf M}_{B}$ amplitude of decays $A^{+}\rightarrow B^{++}l^{-}\overline{\nu}_l \gamma$
is given in terms of the known amplitudes of $A^{0}\rightarrow B^{+}l
^{-}\overline{\nu}_l \gamma$ and of $A^{-}\rightarrow B^{0}l^{-}\overline{\nu}_l \gamma$, which
can be found in Ref.~\cite{c2}.

Let us now show that, as in the virtual RC case, the same linear
combinations of the amplitudes (47) and (48) can be obtained for the
differential decay rates. After squaring and summing over spins and photon
polarization [making the replacement $u_{1}$ $\rightarrow $ $\Sigma ({\not \! s}_{1}) u_{1}$ in
the polarized case], one gets from Eq.~(47)
\begin{eqnarray}
\sum \left| {\sf M}_{B}( A^{++}\rightarrow B^{+}l^{+}\nu_l \gamma )
\right|^{2} & = & 4\sum \left| {\sf M}_{B}( A^{+}\rightarrow B^{0}l^{+}\nu_l
\gamma ) \right|^{2}+\sum \left| {\sf M}_{B} ( A^{0}\rightarrow
B^{-}l^{+}\nu_l \gamma ) \right|^{2} \nonumber \\
&  & \mbox{} -4{\rm Re}\sum {\sf M}_{B} ( A^{0}\rightarrow B^{-}l^{+}\nu_l \gamma)
\overline{{\sf M}}_{B}( A^{+}\rightarrow B^{0}l^{+}\nu_l \gamma ).
\end{eqnarray}

This equation is the same as
\begin{eqnarray}
\sum \left| {\sf M}_{B}( A^{++}\rightarrow B^{+}l^{+}\nu_l \gamma)
\right|^{2} & = & 2\sum \left| {\sf M}_{B}( A^{+}\rightarrow B^{0}l^{+}\nu_l
\gamma ) \right|^{2}-\sum \left| {\sf M}_{B}( A^{0}\rightarrow
B^{-}l^{+}\nu_l \gamma) \right|^{2} \nonumber \\
&  & \mbox{} + 2\sum \left| {\sf M}_{B}( A^{+}\rightarrow B^{0}l^{+}\nu_l \gamma)
-{\sf M}_{B}( A^{0}\rightarrow B^{-}l^{+}\nu_l \gamma) \right|^{2}.
\end{eqnarray}

In the last term on the rhs the contributions of zeroth order in $(q/M_1)$, that is contributions
of order $(\alpha/\pi)(q/M_1)^0$, are the same in both amplitudes
${\sf M}_{B}( A^{+}\rightarrow B^{0}l^{+}\nu_l \gamma)$ and ${\sf M}_{B}( A^{0}\rightarrow 
B^{-}l^{+}\nu_l \gamma)$. Therefore this difference of amplitudes is of order $(\alpha/\pi)(q/M_1)$.
Accordingly, its square is of order $(\alpha/\pi)(q/M_1)^2$ and should be neglected.

Hence, the differential decay rate with bremsstrahlung radiative corrections
has the form
\begin{eqnarray}
d\Gamma_{B}( A^{++}\rightarrow B^{+}l^{+}\nu_l) =2d\Gamma_B( A^{+}\rightarrow B^{0}l^{+}\nu_l)
-d\Gamma_{B}( A^{0}\rightarrow B^{-}l^{+}\nu_l).
\end{eqnarray}

Repeating the same analysis, we obtain for processes $A^{+}\rightarrow
B^{++}l^{-}\overline{\nu}_l$
\begin{eqnarray}
d\Gamma_{B}( A^{+}\rightarrow B^{++}l^{-}\overline{\nu}_l) =2d\Gamma
_{B}( A^{0}\rightarrow B^{+}l^{-}\overline{\nu}_l) -d\Gamma_{B} (A^{-}\rightarrow
B^{0}l^{-}\overline{\nu}_l).
\end{eqnarray}

Adding Eqs.~(35) and (51) for the process $A^{++}\rightarrow B^{+}l
^{+}\nu_l $, and adding Eqs.~(36) and (52) for $A^{+}\rightarrow B^{++}l
^{-}\overline{\nu}_l $, we get the complete differential decay rates with radiative
corrections up to order $(\alpha/\pi)(q/M_1)$,
\begin{eqnarray}
d\Gamma ( A^{++}\rightarrow B^{+}l^{+}\nu_l) =2d\Gamma (
A^{+}\rightarrow B^{0}l^{+}\nu_l ) -d\Gamma( A^{0}\rightarrow B^{-}l^{+}\nu_l),
\end{eqnarray}
\begin{eqnarray}
d\Gamma ( A^{+}\rightarrow B^{++}l^{-}\overline{\nu}_l ) =2d\Gamma (A^{0}\rightarrow B^{+}l^{-}
\overline{\nu}_l) - d\Gamma ( A^{-}\rightarrow B^{0}l^{-}\overline{\nu}_l).
\end{eqnarray}
This completes our study of RC to decays in groups (9) and (10).

\section{Discussion}

BSD with heavy quarks involved present many choices for the charges
of the participating baryons. All these decays can be classified into six
different groups, (5)-(10). The RC to these decays were calculated assuming
the specific charge assignments of groups (5) and (6). To cover the other
assignments one faces the need of recalculating the RC. This can be avoided
by reviewing the previous calculations. In this paper we obtained first the
changes to be made in the final results of the calculations for decays in
groups (5) and (6) to obtain the final results of the charge assignments in
groups (7) and (8) when $A$ is polarized. These changes take the form of a
very practical rule, which complements the rule when $A$ is unpolarized.
Second, we proceeded to determine the changes required in the final results
of decays in groups (5)-(8) to obtain the final results of decays in
groups (9) and (10). The final results for the latter are given as simple
linear combinations of the final results of decays in the four former groups. In
short, the RC to decays in groups (5) and (6) can be directly used to
obtain the final results of all charge assignments of BSD involving heavy
quarks.

Although we studied specifically the case of charm being the heavy quark,
our conclusions also apply if bottom is the heavy quark and if several heavy
quarks, one charm and one bottom, both bottom, etc., are present. The case
of top being one of the heavy quarks would also be covered by the charge
assignments (6)-(10), although it is not expected that top will form bound
baryonic states. This last possibility is only of academic interest.

Our results are model-independent and they are not compromised to particular
values of the form factors. The charged lepton is not restricted in any way,
so $l^{\pm }$ may be $e^\pm$, $\mu^\pm$, or $\tau^\pm$.
Both cases of polarized or unpolarized decaying baryon $A$ are covered. The
three-body and four-body regions of the DP are also covered. For unpolarized
$A$, previous results for (5) and (6) are complete to orders $(\alpha/\pi)(q/M_1)^0$
and $(\alpha/\pi)(q/M_1)$. When $A$ is polarized they are complete to order
$(\alpha/\pi)(q/M_1)^0$. In the near future we hope to include contributions of order
$(\alpha/\pi)(q/M_1)$.

Before closing let us discuss the practical usefulness of our results. For
this purpose we must assess the validity of our approximations with respect
to experimental error bars of observables in BSD. By low, medium, and high
statistics experiments we mean those with several hundreds, thousands, and
hundreds of thousands of events, respectively. The corresponding error bars
of the observables are around 6-10\%, 2-5\%, and 1\% or better,
respectively. Our model-independent RC must be discussed also in percentage.
It is convenient to separate the RC in the Coulomb and in the non-Coulomb
contributions, when the former is present. The reason for this is that it is
appreciably larger than the latter by a factor two and even close to five in some
cases. From previous numerical calculations we know that the non-Coulomb
part is around 1\% and may come close to 2\%. Conservatively, we can then
take this part to be 2\%.

Let us further clarify our approximations. They consisted in neglecting the
model-dependent terms of order $(\alpha/\pi)(q/M_1)^2$ and higher in the RC.
This of course does not mean that we have neglected all terms of such orders
in the model-independent part. In particular, it must be noted that the Coulomb
part arising from the pole in the amplitude of Fig.\ 1a, whose position is
fixed by the velocity $\beta$ of $l^+$, gets its $(q/M_1)^2$ and higher order
corrections from the uncorrected differential decay rate. Therefore all of
these higher order contributions are accounted for. It is then clear that
our approximations apply only to the non-Coulomb part.

The variation of $q/M_1$ in processes (1) to (4) is quite important.
Taking $q/M_1\simeq (M_1-M_2)/M_1$ one can make estimates for this ratio in
different BSD. In (1) $q/M_1\simeq 0.06$, in (2) $q/M_1\simeq 0.20$, in (3)
and (4) $q/M_1\simeq 0.5$. In decays when the heavier bottom quark is involved
this ratio may become even larger. For $b\rightarrow c$ one has $q/M_1\simeq 0.5$
again, but for $b\rightarrow s, u,d$ one has $q/M_1\simeq 0.8$. To make an
estimate of the percentage theoretical error involved in our approximations
it is not convenient to use the product $(\alpha/\pi)(q/M_1)^N$ with $N=2, 3$
because one obtains very small numbers this way. It is better to multiply the
above conservative estimate of 2\% of the non-Coulomb RC [involving the order
$(\alpha/\pi)(q/M_1)^N$ with $N=0, 1$] and multiply it by $(q/M_1)^N$ with
$N=2, 3$. For (1) these numbers are negligibly small. For (2) they are $0.08\%$
and $0.016\%$ for $N=2, 3$, respectively, and are also negligible. For (3) and
(4) they are $0.5\%$ and $0.25\%$, and adding them one has $0.75\%$. The case
$b\rightarrow c$ repeats (3). When $b\rightarrow s, u, d$ one has $1.2\%$ and
$0.9\%$ for $N=2, 3$, respectively. The sum gives an estimate of $2.1\%$. It is
customary that the theoretical error (or bias) be folded in quadratures with the
experimental error bars. One can then ignore this bias if the latter is not
changed appreciably. A change of no more than $10\%$ is reasonable. This means
that the theoretical bias should not exceed one-half of the experimental error
bars. To remain conservative we shall require that the former does not exceed
one-fourth of the latter.

We can draw conclusions comparing the above estimates with the expected
error bars of experiments. When (1) and (2) are involved our results are
very good in high statistics experiments. For (3), (4) and $b\rightarrow c$
they are good in low and still acceptable medium statistics experiments.
However, they are no longer good for high statistics experiments. The same
occurs in the cases $b\rightarrow s, u, d$, they are quite good in
low statistics experiments and are still acceptable in medium statistics
experiments. However, in high statistics ones they are no longer acceptable
in general. It is in these latter cases that terms of order $(\alpha/\pi)
(q/M_1)^2$ and higher should be added to our results. This should be done in
the future when eventually it becomes necessary. Although, then one will have
to face problem of the model-dependence of RC.

However, RC can be rendered smaller in particular cases because of
cancellations due to the values of the form factors involved. This must be
checked case by case. Thus, it may still be possible to use our results in
high statistics experiments when such fortuitous cancellations occur.
Finally, let us stress that our results use previous expressions that are
either analytical or presented in a form ready to be integrated numerically.
Previous numerical results can not be used because masses and form factors
were given particular values.

\acknowledgements

The authors wish to express their gratitude to Consejo Nacional de Ciencia y
Tecnolog{\'\i}a (Mexico) for partial support. Two of us (AMV and
JJT) express their gratitude to Comisi\'on de Operaci\'on y Fomento de
Actividades Acad\'emicas (Instituto Polit\'ecnico Nacional).

\end{document}